\documentclass[12pt]{iopart}
\usepackage{iopams}
\usepackage{tikz}
\tikzset{>=latex}
\usetikzlibrary{arrows}
\usetikzlibrary{automata}
\usepackage{subcaption}
\usepackage{bm}
\usepackage{url}
\usepackage{hyperref}
\usepackage{upgreek}
\usepackage{cleveref}
\usepackage[pagewise]{lineno}

\begin{document}

\title[Source term for ${}^{18}$F production]{Source term calculation and validation for ${}^{18}$F-production with a cyclotron for medical applications at HZDR}

\author{J Konheiser$^1$, S E M\"uller$^1$, A Magin$^{2,a}$, B Naumann$^1$ and A Ferrari$^1$}

\address{$^1$Helmholtz-Zentrum Dresden-Rossendorf, 01328 Dresden,  
  Germany}
\address{$^2$Karlsruhe Institute of Technology (KIT), 76131 Karlsruhe,  
  Germany}
  \address{$^a$now at: SICK AG, 79276 Reute, Germany}
\ead{j.konheiser@hzdr.de}
\vspace{10pt}
\begin{indented}
\item[]May 2018
\end{indented}

\begin{abstract}
In this document we present the calculation and experimental validation of a source term for  ${}^{18}$F-production with a cyclotron for medical applications operating at 18 MeV proton energy and  30 $\upmu$A proton current. The Monte Carlo codes MCNP6 and FLUKA were used for the calculation of the source term. In addition, the radiation field around the $^{18}$O-enriched water target was simulated with the two codes. To validate the radiation field obtained in the simulation, an experimental program has been started using activation samples which are placed close to the water target during an ${}^{18}$F-production run of the cyclotron. After the irradiation, the samples are analyzed and the resulting activation is compared to Monte Carlo calculations of the expected sample activation. We find good agreement between simulations and experimental results, with most calculation to experiment (C/E) ratios well between 0.6 and 1.4. 
\end{abstract}

%
\vspace{2pc}
%
\submitto{\JRP}
%
%
%

\section{Introduction}  
Positron emission tomography (PET) has developed into a standard tool for imaging methods in medicine. The required radionuclides are often produced with the aid of cyclotrons. Depending on the emitter to be produced, different nuclides are bombarded with protons or deuterons that trigger  nuclear reactions. In addition to the desired nuclide, neutron and gamma radiation is also produced during these nuclear reactions. These are the main source of the radioactive dose rate on the outside of the protection buildings and determine the shielding design. In addition, neutron radiation leads to the activation of the construction and building materials, which could be important for the decommissioning of the facilities. Therefore, the correct determination of neutron and gamma source terms is the imperative basic condition for a correct shielding calculation and thus for a sufficient protection of the employees.

Several approaches can be taken to obtain the source term needed for the shielding calculations. In one approach, the source term spectrum is determined using nuclear model programs such as ALICE-91~\cite{alice91}, and subsequently, the corresponding transport calculation (shielding calculation) is done using the obtained spectrum. The determination of the absolute number of emitted neutrons is then carried out on the basis of tabulated activities assuming a full absorption of the beam. For many reactions these are available in tabular form for different proton energies and a standard current~\cite{NDS-NEA}. This approach for the ${}^{18}$F-production is applied in the source term quoted in the supporting documentation for an ACSI TR-FLEX cyclotron~\cite{Sherbrooke}, assuming that the neutron source term originates exclusively from the desired reaction. In~\cite{Sheu} the energy and angular
distributions of the neutron source term were taken from
the double differential data of the nearby reaction ${}^{14}$N(p,n)${}^{14}$O, but also basing the absolute number of emitted neutrons on the production rate of ${}^{18}$F. Often the source term can be traced back to a confidential information from the manufacturer of the cyclotron with little information on how it was obtained~\cite{Bosko1,Bosko2,Facure}. 
\newline 
A different approach which is related to the large progress on radiation transport and reaction codes in recent years consist in the direct calculation of the full neutron and gamma source terms including all contributing reaction channels with the radiation transport codes themselves, often accompanied by comparisons of the simulated results with experimental measurements. Examples for this approach can be found in~\cite{Cruzate, Benavente, Infantino}.  

For the shielding calculations for the new cyclotron with proton beam energies of 24 MeV and 28 MeV at the HZDR~\cite{JRP36}, this second approach was used. Different nuclear models are integrated into the program MCNP6~\cite{mcnp6} to calculate the generation of neutrons in the target.  Likewise, the source term can also be determined with the help of corresponding cross section tables. Both possibilities were used. The source terms calculated in this approach show a large difference respect to the values that were obtained using the approach mentioned above on the basis of tabulated activities. A calculation with the FLUKA~\cite{fluka,fluka2} program gave similar results like MCNP6. These results have been already published and discussed in~\cite{JRP36}. \newline
To validate the results from independent radiation transport codes, in addition to source term calculations with MCNP6 and FLUKA for HZDR's 18 MeV cyclotron\footnote{Cyclone 18/9 model by IBA}, this work gives experimental results for neutron fluence measurements using activation sample monitors. For these measurements, existing experience from the field of reactor dosimetry was applied.

\section{Determination of the neutron source term}
\label{Sec:SourceTerm}
To calculate the neutron source term, simulation models were created with both MCNP6 and FLUKA which consist of a cylinder with radius 0.55 cm and a length of 4 cm, filled with water enriched with 97\% \textsuperscript{18}O. 
These dimensions correspond to typical target bodies used at the IBA cyclotron. 
The protons' direction is  along the cylinder axis, hitting the target on one of the circular base surfaces. The precise shape of the proton beam is not known. Therefore, two approaches were calculated. In the first case the proton beam was simulated as an infinitesimally small pointlike beam and in the second case a circular surface beam with a Gaussian distribution with a standard deviation of 0.125 cm cut off at the target radius was chosen. The two approaches gave identical results (see also~\cite{JRP36}). In the following we will use the results obtained with a pointlike proton beam in the simulations. The emitted neutron spectrum is determined on the surface of a surrounding sphere with a radius of 10 m, large enough to minimize geometrical effects due to the target shape. 
The generation of neutrons in the target was carried out using nuclear physics models of reaction cross sections. In MCNP a cascade exciton model (CEM)~\cite{cem} was used, while FLUKA uses a pre-equilibrium cascade model (PEANUT)~\cite{peanut} for the nuclear interactions. In addition, MCNP6 calculations were also carried out with evaluated nuclear data of the (p,n) reaction. \newline
Since \textsuperscript{18}O-data are not included in the standard library of MCNP6, they were generated using the NJOY~\cite{njoy} program and imported into MCNP6. The required reaction cross-sections were read from the nuclear data library TENDL, which is based on the nuclear core model code TALYS~\cite{talys}. This possibility to use externally generated  cross sections does not exist for FLUKA. Since FLUKA does not include neutron cross sections for \textsuperscript{18}O, cross section data for \textsuperscript{16}O was used instead for the interactions of neutrons in the water. At thermal energies, the O${}^{16}$ total neutron cross section is at maximum 25\% higher than the one for O${}^{18}$. In both cases, the cross sections are dominated by the ones for elastic scattering of neutrons. Since the main source of neutrons is the dominating (p,n) reaction at the O${}^{18}$, the influence of the secondary (n,g), (n,n') and (n,2n) reactions is very small. Further differences in the calculations consist of the used cross section data libraries. The data libraries ENDF/B-VI.8~\cite{bvi8} were used  for the interactions of neutrons with energies below 20 MeV at FLUKA and  ENDF/BVII.1~\cite{bvii1} for MCNP6.
\begin{figure}[t!]
\centering
\begin{subfigure}{0.48\linewidth}
\begin{center}
    \includegraphics[trim=25 10 0 30 , clip, width=0.99\linewidth]{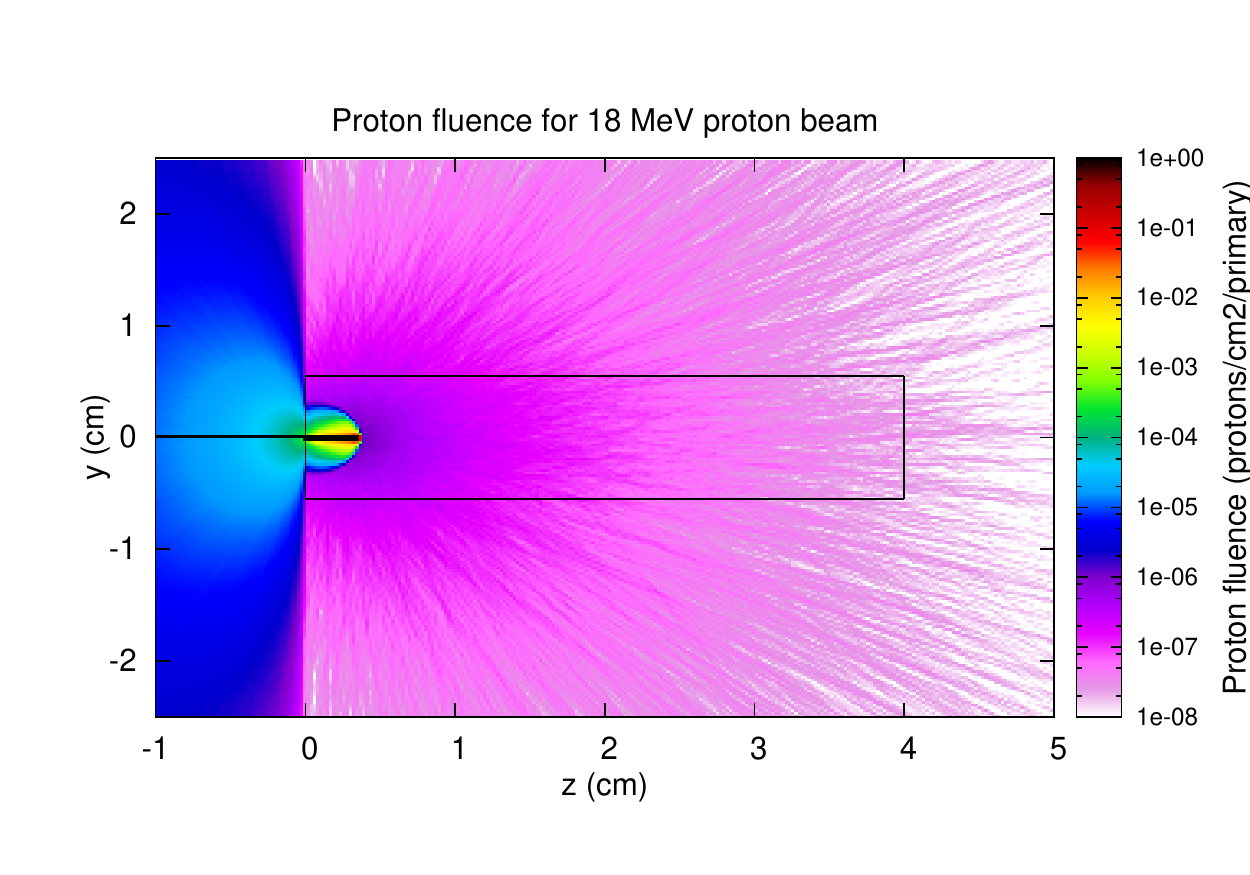}
\end{center}
\end{subfigure}
\hspace{0.2cm}
\begin{subfigure}{0.48\linewidth}
\begin{center}
    \includegraphics[trim=25 10 0 30 , clip, width=0.99\linewidth]{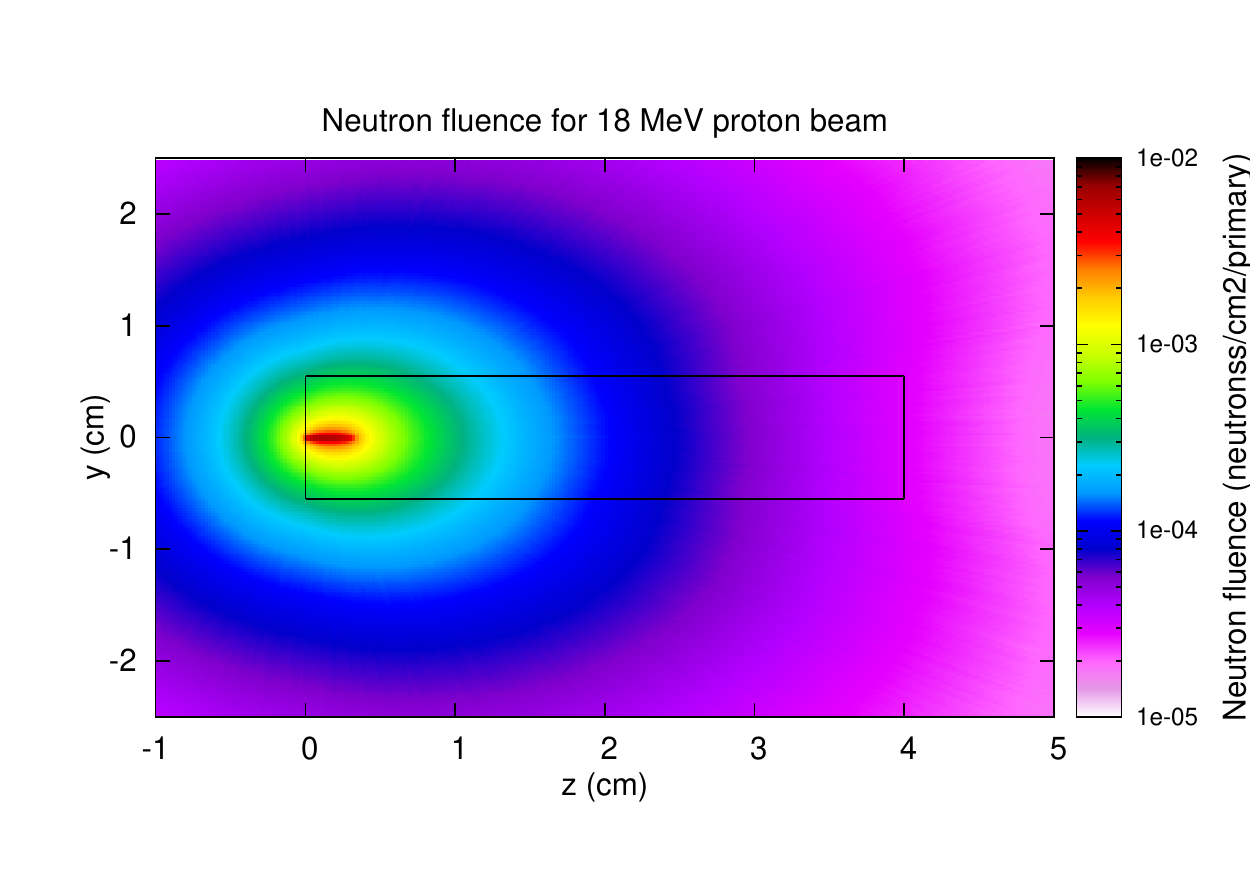}
\end{center}
\end{subfigure}
\caption{Proton- and neutron fluences per primary proton obtained with FLUKA for 18 MeV protons hitting the \textsuperscript{18}O-enriched water target.}\label{Fig:Fluence18MeV}
\end{figure}
In fig.~\ref{Fig:Fluence18MeV}, both the proton and neutron fluences per primary proton are depicted as obtained with the FLUKA simulation code. The protons penetrate only about 0.5 cm into the water target before they are stopped. The water target absorbs almost all protons in the forward direction, leaving only the backscattered ones to the left of the target. Neutrons are produced along the trajectory of the proton beam in the water.

\begin{figure}[t!]
\centering
\includegraphics[trim=10 100 10 70 , clip, width=0.99\linewidth]{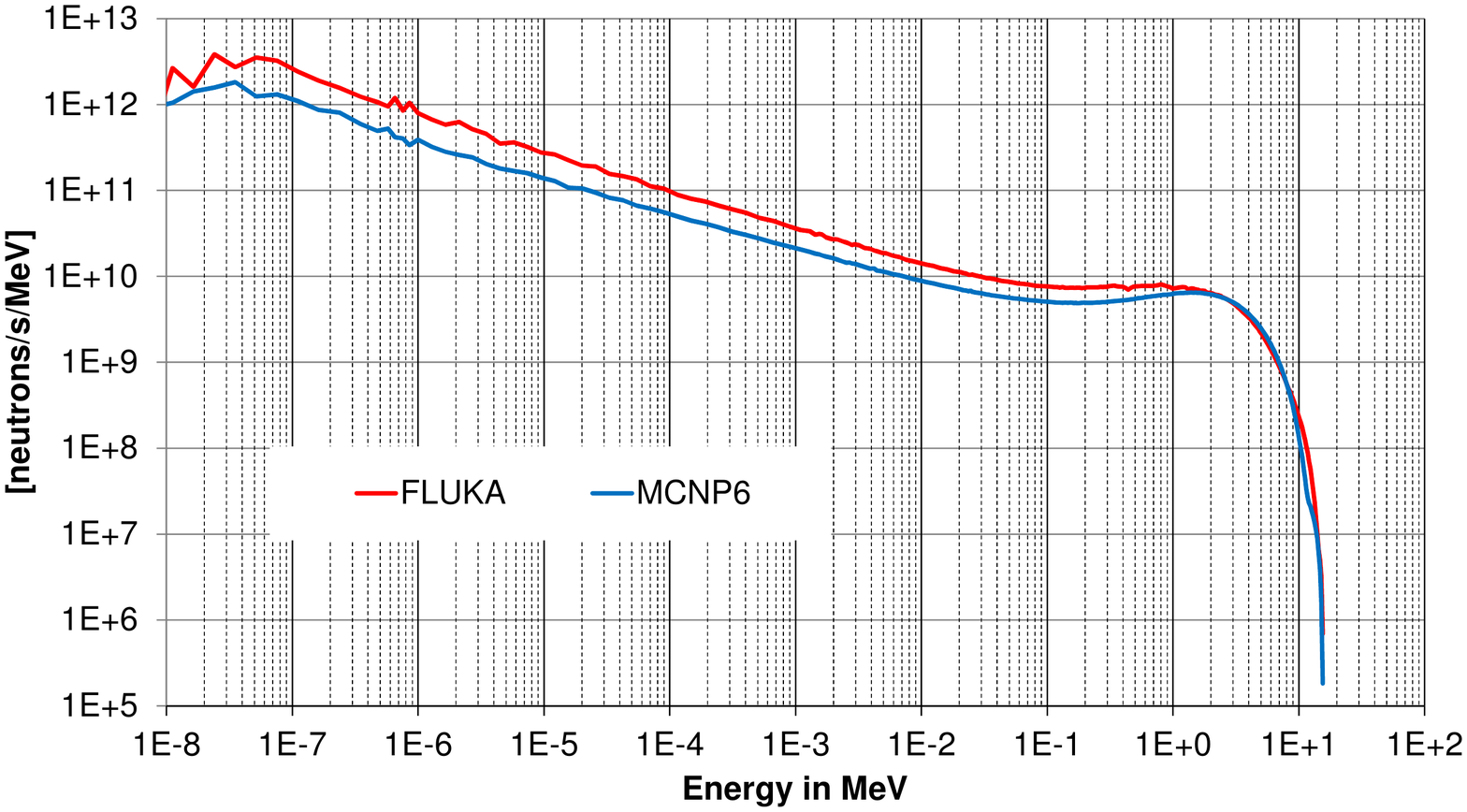}
\caption{Differential neutron rate for an 18 MeV proton beam with 1$\upmu$A hitting the water target. The spectra are available from~\cite{data_Fig2}.}
\label{Fig:nRate}
\end{figure}

Fig.~\ref{Fig:nRate} shows the differential neutron rate recorded across the surrounding sphere for 1$\upmu$A of proton beam current obtained with MCNP6 (version 6.1.1) and FLUKA (version 2011.2x). Integrating the spectrum over energy we find a total neutron yield of $3.21\times10^{10}$ n/s for 1 $\upmu$A of proton current for the FLUKA calculation, and $2.99\times10^{10}$ n/s for 1 $\upmu$A of proton current for MCNP6. The higher yield obtained with FLUKA respect to the MCNP6 calculations has already been observed for 24 and 28 MeV protons in ~\cite{JRP36}, and is attributed to differences of the underlying nuclear physics models. The values are about a factor 3 higher than the value of 1.115$\times$10\textsuperscript{10} n/s for 1 $\upmu$A of proton current obtained from~\cite{IAEA} for the ${}^{18}$O(p,n)${}^{18}$F channel. We attribute the difference to additional neutron-producing reaction channels opening at 18 MeV proton energy, as suggested in~\cite{Carroll}. It should be noted however that measurements of the neutron yield rate reported in~\cite{Mendez} and~\cite{Hagiwara} give results which are close to the value of 1.115$\times$10\textsuperscript{10} n/s for 1 $\upmu$A\footnote{It was confirmed by the authors of~\cite{Hagiwara} that the values quoted in their document need to be corrected by a factor of 10 and the resulting neutron production yield should therefore read  (1.55$\times$10\textsuperscript{10}$\pm$1.03$\times$10\textsuperscript{9}) n/s for 1 $\upmu$A of beam current. We thank M. Hagiwara for this information.}.

\section{Experimental validation of the radiation field around the target}  
To validate the calculation of the source terms in sec.~\ref{Sec:SourceTerm}, activation monitor foils where placed on top of the irradiation target during a routine run for \textsuperscript{18}F production. After irradiation, the activation of the foils was measured and compared to predictions from the radiation transport and reaction codes MCNP6 (version 6.1.1) and FLUKA. For these activation studies, a special developer version of FLUKA~\cite{FLUKA2017} was used which includes updated information on branching ratios to meta-stable states using the JEFF-3.1A activation library~\cite{JEFF31A} that is not yet available in the official FLUKA version~\cite{FLUKA2011}.     

\subsection{Experimental setup to measure the radiation field with sample activation}
\begin{figure}[th!]
\begin{subfigure}{0.48\linewidth}
\begin{tikzpicture}[>=stealth]
  \node[anchor=south west, inner sep=0] (image) at (0,0){
\includegraphics[trim=0 240 0 0 , clip, width=0.95\linewidth]{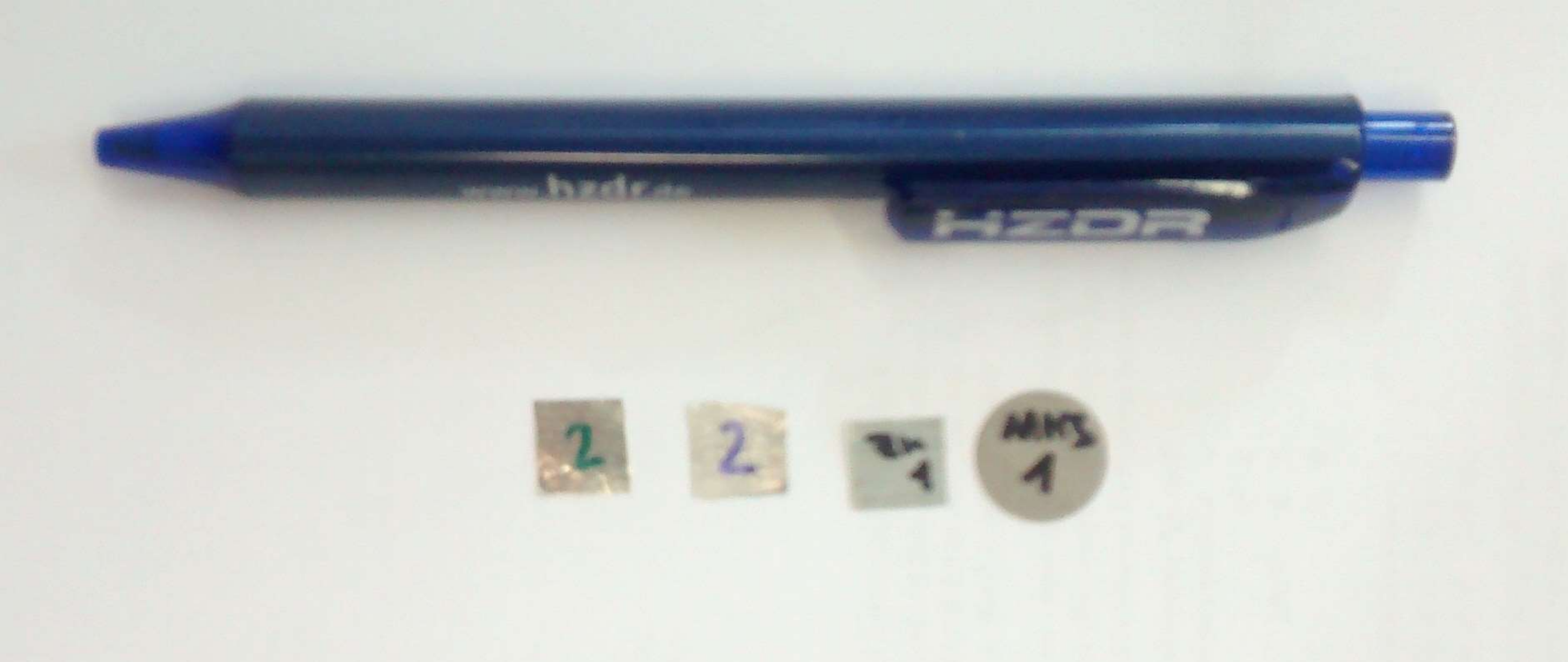}};
       \begin{scope}[x={(image.south east)},y={(image.north west)}]
         \draw (0.1,0.58) node [anchor=north west,very thick] {\bf \sffamily Tin};
         \draw (0.08,0.45) node [anchor=north west,very thick] {\tiny \bf \sffamily 100.\% Sn};
         \draw[black,very thick] (0.34,0.335) -- (0.22,0.45);         
         \draw (0.4,0.225) node [anchor=north,very thick] { \bf \sffamily Indium};
         \draw (0.4,0.12) node [anchor=north,very thick] {\tiny \bf \sffamily 99.9\% In};         
         \draw[black,very thick] (0.42,0.18) -- (0.47,0.26);
         \draw (0.65,0.225) node [anchor=north,very thick] { \bf \sffamily Zinc};
         \draw (0.65,0.12) node [anchor=north,very thick] {\tiny \bf \sffamily 100.\% Zn};         
         \draw[black,very thick] (0.57,0.24) -- (0.6,0.17);
         \draw (0.75,0.56) node [very thick] {\bf \sffamily Multi-component};
         \draw (0.87,0.52) node [anchor=north,very thick] {\tiny \bf \sffamily 81.63\% Ni};
         \draw (0.87,0.45) node [anchor=north,very thick] {\tiny \bf \sffamily 15.16\% Mo};
         \draw (0.87,0.38) node [anchor=north,very thick] {\tiny \bf  \sffamily 2.62\% W};
         \draw (0.87,0.31) node [anchor=north,very thick] {\tiny \bf  \sffamily 0.26\% Mn};
         \draw (0.87,0.24) node [anchor=north,very thick] {\tiny \bf \sffamily  0.31\% Au};         
    \draw[black,very thick] (0.75,0.52) -- (0.684,0.4);    
       \end{scope}
\end{tikzpicture}
\caption{}\label{Fig:Samples_a}
\end{subfigure}
\begin{subfigure}{0.48\linewidth}
\centering
\includegraphics[trim=0 0 200 0 , clip, width=0.95\linewidth]{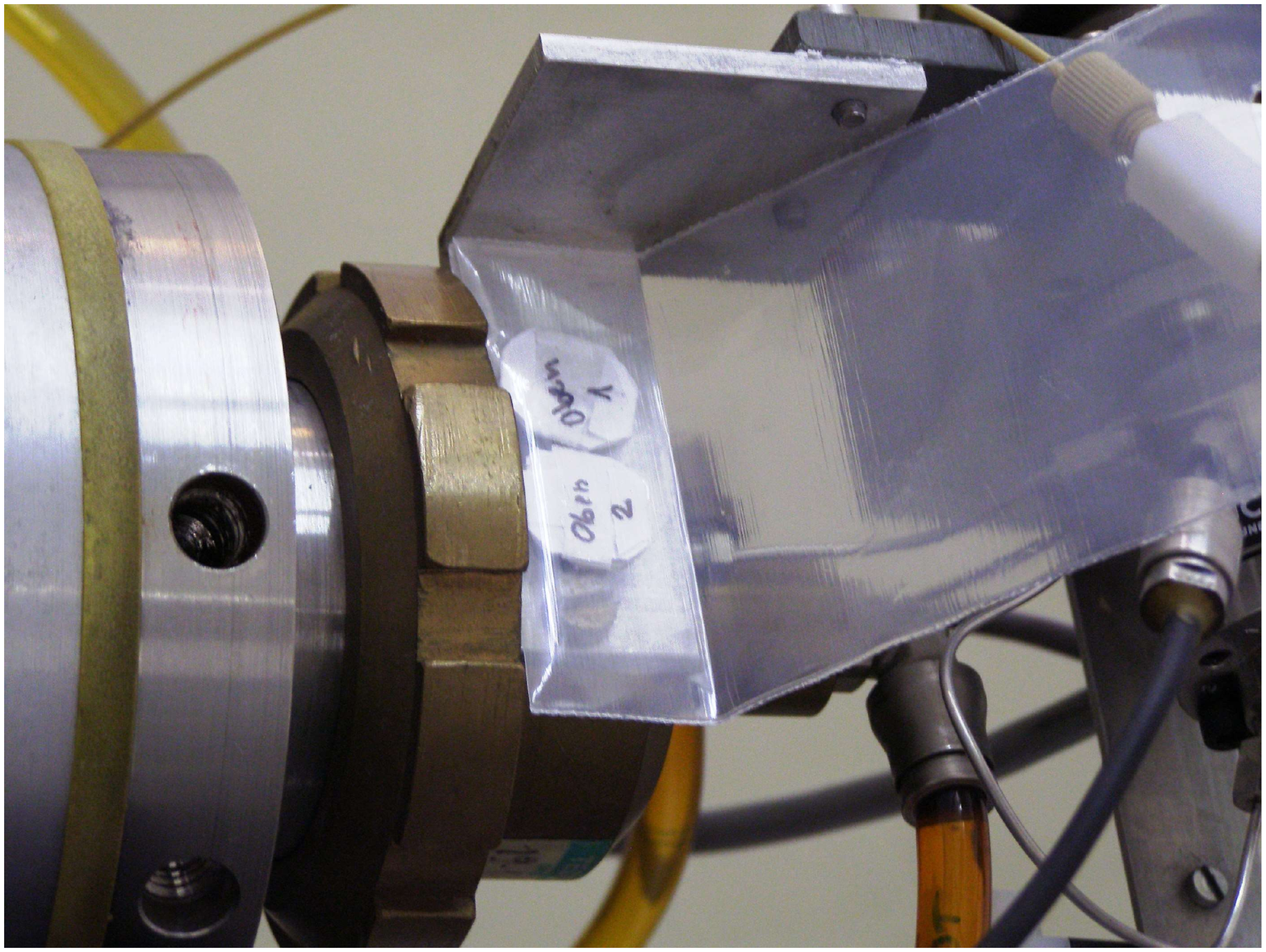}
\caption{}\label{Fig:Samples_b}
\end{subfigure}
\caption{(a) Examples for activation foil samples used in the experiment. (b) The two stacks of activation sample foils in a plastic bag placed on top of the target flange.}\label{Fig:Samples}
\end{figure}

Figures~\ref{Fig:Samples_a} and~\ref{Fig:Samples_b} show the individual activation monitor samples as well as the sample packages and their position on the irradiation container at the cyclotron. The samples consist of different metal foils made of pure metals or alloys. Table~\ref{Tab:1} shows the monitor samples and the reactions under study with the generated nuclides, the reaction threshold and their half-life. The selected metals are standard monitor materials which are inserted for neutron flux and fluence measurements at fission reactors for power determinations as well as for the validation of the results in reactor dosimetry~\cite{dosimetry}. As can be seen from table~\ref{Tab:1}, several of the materials have reactions starting at different threshold energies. This makes it possible to study different energy regions in the spectrum. The monitor packages were positioned directly on top of the irradiation target in order to achieve a high neutron flux and thus high reaction rates. The irradiation took place during a regular ${}^{18}$F production run. The energy of the protons was 18 MeV with average beam current of 25 $\upmu$A and the irradiation lasted for 50 minutes.

\begin{table}[t!]
\caption{Composition of the monitor samples with the studied activation reactions and corresponding half-lifes.}
\centering
\begin{tabular}{|c|c|c|c|r|}
\hline
Monitor Sample & Mass fraction & Reactions & Threshold & Half-life\\
\hline
Multi-component & 81.63\% Ni& \textsuperscript{58}Ni(n,np)\textsuperscript{57}Co      & 8 MeV & 271.74d\\
 &            & \textsuperscript{58}Ni(n,p)\textsuperscript{58}Co & 0.4 MeV &  70.86d\\
 & 15.16\% Mo & \textsuperscript{98}Mo(n,g)\textsuperscript{99}Mo &  therm. & 66.0h\\
 &            & \textsuperscript{100}Mo(n,2n)\textsuperscript{99}Mo &  8 MeV     &   66.0h   \\ 
 &  2.62\% W  & \textsuperscript{186}W(n,g)\textsuperscript{187}W   & therm.  & 23.72h\\
 &  0.26\% Mn & \textsuperscript{55}Mn(n,g)\textsuperscript{56}Mn   &  therm. &   2.58h\\ 
 &  0.31\% Au & \textsuperscript{197}Au(n,g)\textsuperscript{198}Au & therm. (4eV)  &   2.69d\\
\hline
Zinc & 100.00\% Zn& \textsuperscript{64}Zn(n,p)\textsuperscript{64}Cu  &0.08 MeV  & 12.70h\\
   &              & \textsuperscript{64}Zn(n,g)\textsuperscript{65}Zn  & therm.  & 244d \\
   &              & \textsuperscript{68}Zn(n,g)\textsuperscript{69m}Zn & therm.  & 13.76h \\
\hline
Indium & 99.9\% In& \textsuperscript{113}In(n,g)\textsuperscript{114m}In  & therm. & 49.5d \\
  &              &  \textsuperscript{115}In(n,2n)\textsuperscript{114m}In & 9 MeV & 49.5d\\
  &              &  \textsuperscript{115}In(n,n')\textsuperscript{115m}In  & 0.3 MeV & 4.5h \\  
\hline
Tin & 100.00\% Sn& \textsuperscript{116}Sn(n,g)\textsuperscript{117m}Sn & therm. & 13.60d\\
    &            & \textsuperscript{117}Sn(n,n')\textsuperscript{117m}Sn & 0.15 MeV & 13.60d\\ 
    &            & \textsuperscript{118}Sn(n,2n)\textsuperscript{117m}Sn & 9 MeV & 13.60d\\ 
\hline
\end{tabular}
\label{Tab:1}
\end{table}
Two packages with the same stacks of activation monitors were irradiated simultaneously. This allowed independent measurements using two independent laboratories for the sample activation analysis. The activation measurements were carried out at the "Department of Environmental Monitoring" and at the "Laboratory for Environment and Radionuclide Analysis" of the "VKTA - Strahlenschutz, Analytik \& Entsorgung Rossendorf e. V."\footnote{VKTA - Radiation Protection, Analytics \& Disposal Inc., \url{www.vkta.de}}. The activity was determined by gamma spectrometry using high-purity germanium detectors. Typically, the detectors\footnote{Broad Energy HPGe Detector BE5030P by Mirion Technologies, Inc.} have about 44\% efficiency relative to a $3\,"\times 3\,"$ NaI(Tl) detector, a resolution of 1.78~keV at 1332~keV and are calibrated weekly against a certified standard. 
In order to detect nuclides with a relatively short half-life, some of the activation monitors were already examined about one hour after the end of irradiation. For some nuclides with longer half-life, the measurements were repeated at longer cooling times and the activity value at the end of irradiation was extrapolated back using the known half-life values. However, it was found in previous studies that these extrapolations were not always reliable, especially when there is a delayed production of a nuclide from an excited state. Therefore both labs were asked to provide the measured values at the time of the measurement and also the values extrapolated back to the time at end of irradiation (EOI). The quoted uncertainties by both laboraties were in the range between 5\% and 25\%, depending on the reaction channel. 

\subsection{Calculation of sample activation}
\label{CalcSampAct}
\begin{figure}[th!]
\centering
\begin{tikzpicture}
\node[anchor=south west, inner sep=0] (image) at (0,0) {
\begin{subfigure}{0.35\linewidth}
\begin{center}
    \includegraphics[trim=77.7 0 0 0 , clip, width=1.40\linewidth]{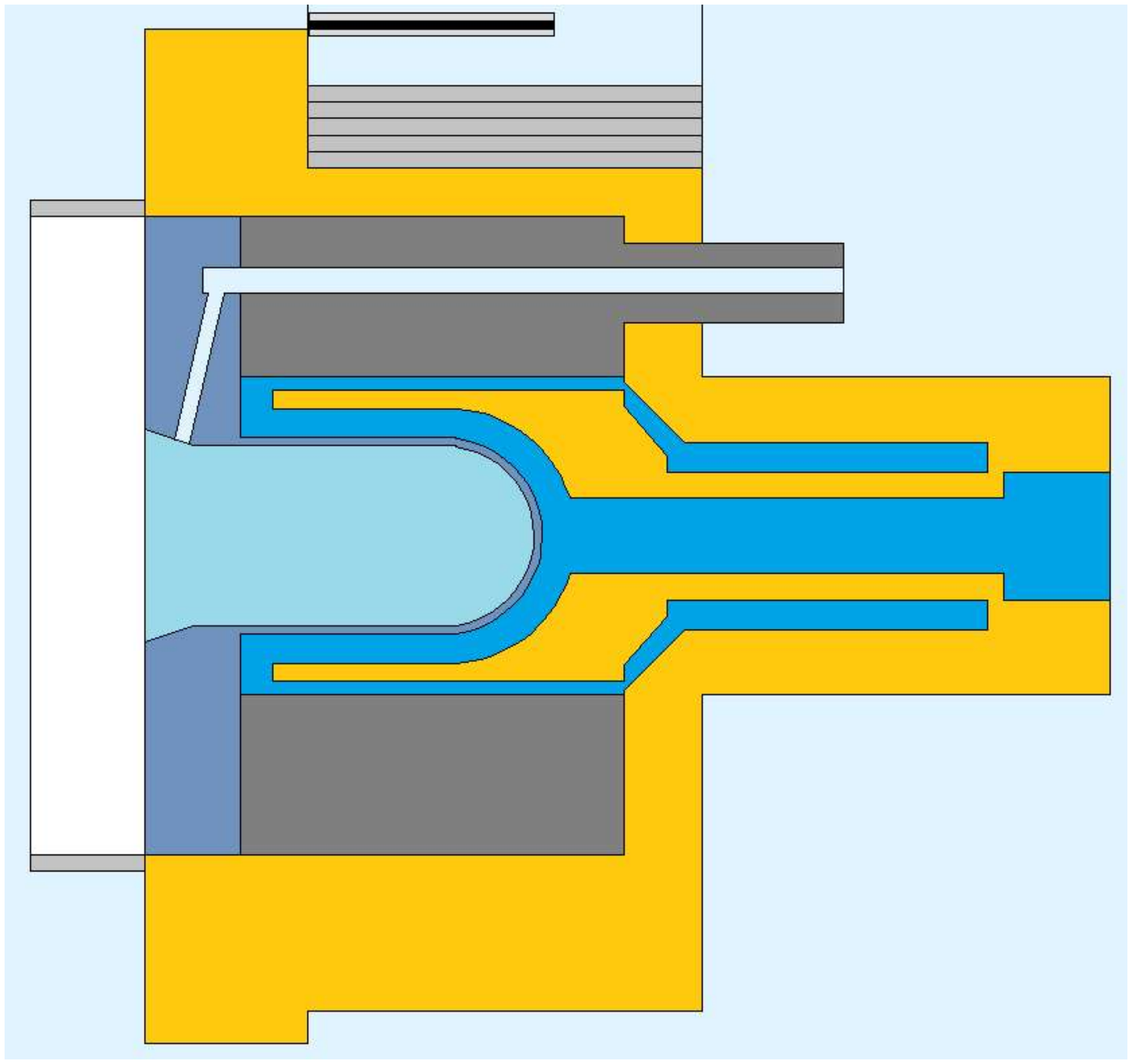}
    \caption{FLUKA geometry model}
\end{center}
\end{subfigure}
\hspace{2.5cm}
\begin{subfigure}{0.38\linewidth}
\begin{center}
    \includegraphics[trim=0 0 0 0 , clip, width=1.4\linewidth]{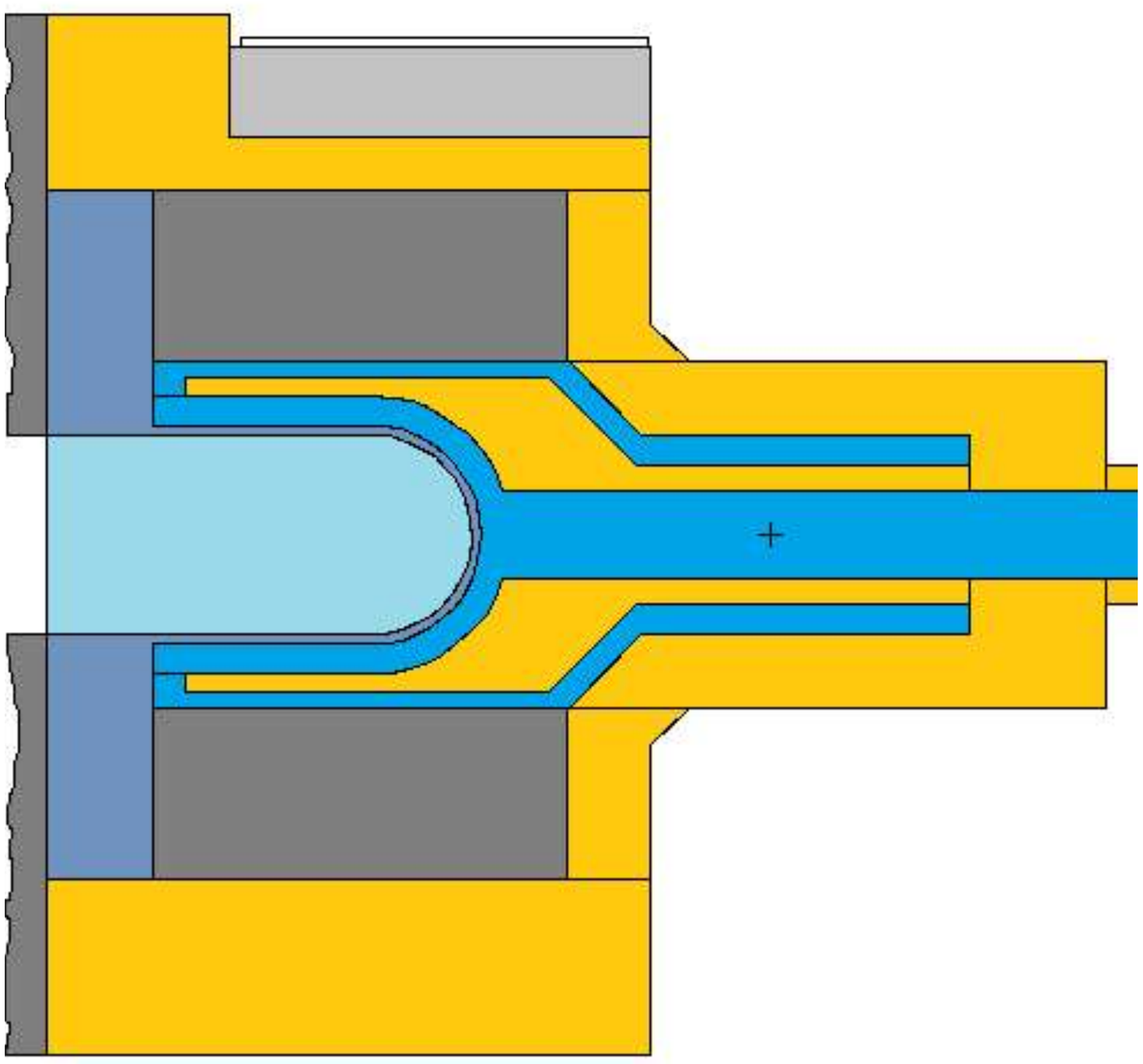}
    \caption{MCNP6 geometry model}
\end{center}
\end{subfigure}
};
\begin{scope}[x={(image.south east)},y={(image.north west)}]
\draw[->,black,ultra thick] (-0.075,0.545) -- (0.005,0.545);
\draw (-0.035,0.54) node [anchor=north] {{\bf  \sffamily proton}};
\draw (-0.035,0.48) node [anchor=north] {{\bf  \sffamily beam}};

\draw (0.47,1.) node [anchor=south] (actsamp) {{\bf  \sffamily activation samples}};
\draw (0.49,.9) node [anchor=south] (brass) {{\bf  \sffamily brass body}};
\draw (0.48,.8) node [anchor=south] (cool) {{\bf  \sffamily cooling water}};
\draw (0.48,.7) node [anchor=south] (targ) {{\bf  \sffamily target (H${}_{\bm{\mathsf 2}}{}^{\bm{\mathsf 18}}$O)}};

\draw (0.49,.3) node [anchor=north] (niob) {{\bf  \sffamily niobium}};
\draw (0.49,.2) node [anchor=north] (steel) {{\bf  \sffamily stainless steel}};

\draw[-{>[scale=.8]}, black,  thick] (actsamp.west) -- (0.165, 0.94);
\draw[-{>[scale=.8]}, black,  thick] (actsamp.east) -- (0.75, 0.9);
\draw[-{>[scale=.8]}, black,  thick] (brass.west) -- (0.2, 0.80);
\draw[-{>[scale=.8]}, black,  thick] (brass.east) -- (0.65, 0.85);
\draw[-{>[scale=.8]}, black,  thick] (cool.west) -- (0.1515, 0.62);
\draw[-{>[scale=.8]}, black,  thick] (cool.east) -- (0.72, 0.638);
\draw[-{>[scale=.8]}, black,  thick] (targ.west) -- (0.13, 0.545);
\draw[-{>[scale=.8]}, black,  thick] (targ.east) -- (0.7, 0.545);
\draw[-{>[scale=.8]}, black,  thick] (niob.west) -- (0.14, 0.48);
\draw[-{>[scale=.8]}, black,  thick] (niob.east) -- (0.67, 0.455);
\draw[-{>[scale=.8]}, black,  thick] (steel.west) -- (0.12, 0.33);
\draw[-{>[scale=.8]}, black,  thick] (steel.east) -- (0.685, 0.33);
\draw[->,black,ultra thick] (0.52,0.545) -- (0.59,0.545);
\draw (0.55,0.54) node [anchor=north] {{\bf  \sffamily proton}};
\draw (0.55,0.48) node [anchor=north] {{\bf  \sffamily beam}};
\end{scope}
\end{tikzpicture}
\caption{Target geometries for the two simulation codes.}\label{Fig:Geometry}
\end{figure}

Fig.~\ref{Fig:Geometry} shows the geometrical models of the irradiation target chamber created with FLUKA and MCNP6. The construction of the irradiation target was reproduced in great detail. While it is known from which materials the components are made of, precise information on densities and composition was not always available, so standard specifications had to be used. Composition and densities of the materials were implemented identically in the FLUKA and MCNP6 simulations. The beam tube adapter flange and surrounding environment like walls were not implemented in the simulations, on the assumption that the influence of back scattered neutrons is negligible on the activation of the monitors. 

\begin{figure}[th!]
\centering
\includegraphics[trim=10 20 10 50 , clip, width=0.85\linewidth]{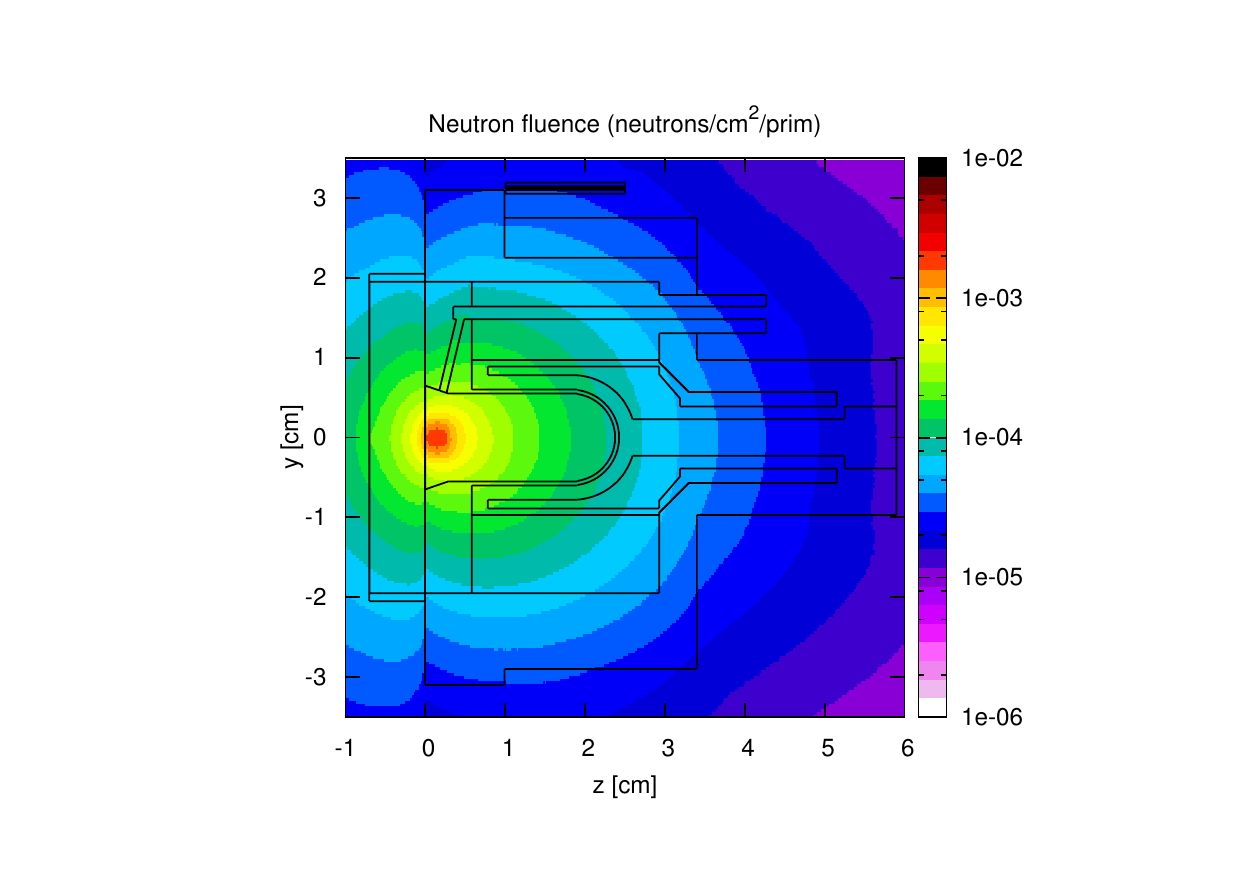}
\caption{Neutron fluence in neutrons/cm${}^2$/primary proton around the target geometry (evaluated with the FLUKA Monte Carlo code).}\label{Fig:NeuFlu}
\end{figure}

The stacks of foil samples were included in the simulations at their corresponding position during the irradiation. Material densities of the samples where measured for each sample before irradiation, and the average value of the samples of same type was used in the simulation. The source term in the simulations consisted of a point-like proton beam with 18 MeV kinetic energy. 
Fig.~\ref{Fig:NeuFlu}  shows the neutron fluence on a central vertical section plane as obtained from the output of a FLUKA simulation. One can clearly see the proton stopping peak immediately after the protons enter the \textsuperscript{18}O-enriched water and the almost isotropic emission of the neutrons through the target geometry.

\begin{figure}[th!]
\centering
\includegraphics[trim=50 60 50 60 , clip, width=0.85\linewidth]{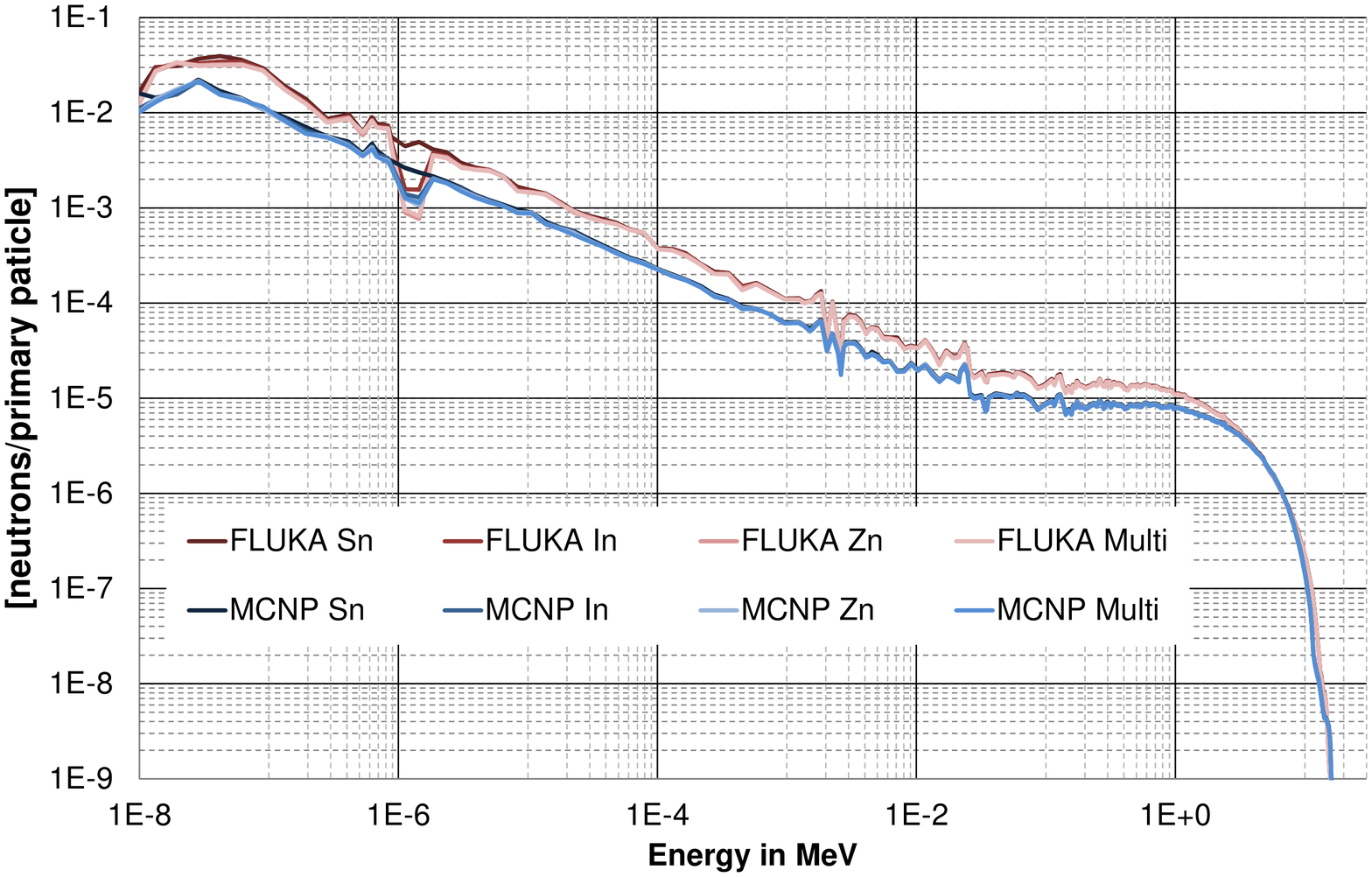}
\caption{Differential neutron flux per primary proton evaluated with FLUKA and MCNP6}\label{Fig:neufluSamp}
\end{figure}

Fig.~\ref{Fig:neufluSamp} shows the spectra of neutron flux entering the different samples as calculated by MCNP6 and FLUKA. Similar to fig.~\ref{Fig:nRate} in sec.~\ref{Sec:SourceTerm}, at energies below 1 MeV, the FLUKA values are higher than the MCNP6 values. The strong  resonance  (at about 1.45 eV) of  the absorption cross section for \textsuperscript{115}In with almost 30000 barn is visible in the spectra, except for tin, because the tin sample was placed below the indium sample. 

Given the neutron flux rate,  the activities $A_i(t_{\mathrm meas})$ for each produced nuclide at a time $t_{\mathrm meas}$ after irradiation in an energy bin $i$  can be determined using the relation
 \begin{equation}
    A_i(t_{\mathrm meas}) = \varrho \cdot V  \cdot \lambda \cdot \sigma_i \cdot \dot{\Phi}_i\cdot t_{\mathrm irr} \cdot   \left(1 - e^{-\lambda t_{\mathrm irr}}  \right) \cdot e^{-\lambda (t_{\mathrm meas})}
\label{Eq:1}    
  \end{equation}  
In eq.~\ref{Eq:1}, $\varrho$ is the density of nuclei in the sample (in nuclei/(barn$\cdot$cm)), $V$ is the sample volume in cm\textsuperscript{3}, $\sigma_i$ is the corresponding reaction cross section in barn for energy bin $i$, $\dot{\Phi}_i$ is the corresponding neutron flux rate obtained from the simulation in neutrons/cm\textsuperscript{2}/s,  $t_{\mathrm irr}$ is the irradiation time in seconds and $\lambda$ is the decay constant of the reaction product (in 1/s). The total activity is then the sum of the $A_i$ over all energy bins $i$. 
Given an irradiation time profile, FLUKA conveniently gives the resulting nuclide activities in Bq/cm\textsuperscript{3} for selected geometry regions at desired times directly in a tabular output. The required cross section data is hard-coded into the FLUKA program and cannot be changed by the user. For MCNP6, eq.~\ref{Eq:1} needs to be applied externally to the simulated neutron flux rates. In this case, the required cross section data had to be generated with the NJOY program. This procedure has the advantage that the neutron flux can be folded over cross sections from different nuclear data libraries. This allows to estimate systematic uncertainties coming from differences between the available cross section data sets. If more than one reaction channel contributed to a measured final state isotope, the resulting activities were added to obtain the final result.

\subsection{Discussion of the results}

In \cref{Tab:2,Tab:3,Tab:4,Tab:5} the results for the measured and simulated activities for the different monitors at the corresponding time of measurement are presented. Measurements obtained by the "Department of Environmental Monitoring" are reported as "Analysis A" and the ones by the "Laboratory for Environment and Radionuclide Analysis" are reported as "Analysis B". We have only kept results for reactions for which both laboratories reported a significant value and for which the statistical uncertainty of the simulations was 15\% or better. This e.g. excludes the reaction \textsuperscript{55}Mn(n,g)\textsuperscript{56}Mn in table~\ref{Tab:1}, for which only Analysis A gave a measured result. In total 11 measurements for different nuclides remain. Numbers in parenthesis correspond to uncertainties on the last digits. The MCNP6 results in the tables use cross section data from the JEFF3.1A libraries for the Indium, Zinc and Tin monitors, while for the multi-component monitor ENDF/B-VII.1 were used (except for the \textsuperscript{58}Ni(n,p)\textsuperscript{58}Co reaction, for which also JEFF3.1A libraries were used).
\begin{table}[t!]
\caption{Measured and calculated activities for the multi-component monitor. The measurements for analysis A were done at $t_{\mathrm meas}=$6h17m after EOI, measurements for analysis B were done at $t_{\mathrm meas}=$30h13m after EOI.  The uncertainty of the FLUKA values corresponds to the statistical uncertainty. A starred value (*) indicates a C/E ratio which is outside the interval [0.6; 1.4].}
\centering
\footnotesize
\begin{tabular}{|c|c|c|c|c|c|c|c|}
\hline
{\bf Multi-comp.} &  \multicolumn{2}{|c|}{Measured activity } & \multicolumn{2}{|c|}{Simulated activity } & \multicolumn{3}{|c|}{Comparison }\\
{\bf monitor}  & \multicolumn{2}{|c|}{[Bq]} & \multicolumn{2}{|c|}{[Bq]} &\multicolumn{3}{|c|}{C/E} \\
\hline
\hline
Reaction &Analysis & Result & MCNP6 & FLUKA & MCNP6 & FLUKA & C\textsubscript{M}/C\textsubscript{F} \\
\hline
\textsuperscript{58}Ni(n,np)\textsuperscript{57}Co & A & 4.2(5) & 0.58 & 0.74(3) & 0.14* & 0.18* &0.78 \\
 & B & 4.7(5) & 0.58 & 0.74(3) & 0.12* & 0.16* & 0.78\\
\hline
 \textsuperscript{58}Ni(n,p)\textsuperscript{58}Co & A & 330(20) & 274 & 255(1) & 0.83 & 0.77 &1.07 \\
  & B & 455(45) & 378 & 351(2) & 0.83 & 0.77 & 1.08 \\
 \hline
 \textsuperscript{99}Mo prod.& A & 120(8) & 72 & 118(3) & 0.60 & 0.98 & 0.61 \\
 & B & 117(12) & 56 & 91.6(2) & 0.48* & 0.78 & 0.61\\
 \hline
 \textsuperscript{186}W(n,g)\textsuperscript{187}W & A & 200(12) & 133 & 239(8) & 0.67 & 1.20 &0.56* \\
 & B & 106(11) & 67 & 120(4) & 0.63 & 1.13 & 0.56*\\
\hline
 \textsuperscript{197}Au(n,g)\textsuperscript{198}Au & A & 68(4) & 41 & 69(4) & 0.60 & 1.02 & 0.59* \\
 & B & 63(6) & 31.4 & 53(3) & 0.50* & 0.72 & 0.59*\\
\hline
\end{tabular}
\label{Tab:2}
\end{table}

\begin{table}[t!]
\caption{Measured and calculated activities for the indium monitor. The measurements for analysis A were done at $t_{\mathrm meas}=$7h47m after EOI, measurements for analysis B were done at $t_{\mathrm meas}=$27h32m after EOI. The uncertainty of the FLUKA values corresponds to the statistical uncertainty. A starred value (*) indicates a C/E ratio which is outside the interval [0.6; 1.4].}
\centering
\footnotesize
\begin{tabular}{|c|c|c|c|c|c|c|c|}
\hline
{\bf Indium} &  \multicolumn{2}{|c|}{Measured activity } & \multicolumn{2}{|c|}{Simulated activity } & \multicolumn{3}{|c|}{Comparison }\\
  & \multicolumn{2}{|c|}{[Bq]} & \multicolumn{2}{|c|}{[Bq]} &\multicolumn{3}{|c|}{C/E} \\
\hline
Reaction & Analysis & Result & MCNP6 & FLUKA & MCNP6 & FLUKA & C\textsubscript{M}/C\textsubscript{F} \\
\hline
\hline
 \textsuperscript{114m}In prod. & A & 29(7) & 7 & 9.9(1) & 0.24* & 0.34* & 0.71 \\
 & B & 25(3) & 6.9 & 9.7(1) & 0.28* & 0.39* & 0.71\\
 \hline
\textsuperscript{115}In(n,n')\textsuperscript{115m}In & A & 9300(465) & 9145 & 9855(41) & 0.98 & 1.06 &0.93 \\
 & B & 474(47) & 434 & 466(2) & 0.92 & 0.98 & 0.93\\
 \hline
\end{tabular}
\label{Tab:3}
\end{table}

\begin{table}[t!]
\caption{Measured and calculated activities for the tin monitor. The measurements for analysis A were done at $t_{\mathrm meas}=$26h43m after EOI, measurements for analysis B were done at $t_{\mathrm meas}=$52h35m after EOI. The uncertainty of the FLUKA values corresponds to the statistical uncertainty. A starred value (*) indicates a C/E ratio which is outside the interval [0.6; 1.4].}
\centering
\footnotesize
\begin{tabular}{|c|c|c|c|c|c|c|c|}
\hline
{\bf Tin} &  \multicolumn{2}{|c|}{Measured activity } & \multicolumn{2}{|c|}{Simulated activity } & \multicolumn{3}{|c|}{Comparison }\\
 & \multicolumn{2}{|c|}{[Bq]} & \multicolumn{2}{|c|}{[Bq]} &\multicolumn{3}{|c|}{C/E} \\
\hline
Reaction & Analysis & Result & MCNP6 & FLUKA & MCNP6 & FLUKA & C\textsubscript{M}/C\textsubscript{F} \\
\hline
\hline
 \hline
\textsuperscript{117m}Sn prod. & A & 84(5) & 55.4 & 66.9(9) & 0.66 & 0.80 &0.83 \\
 & B & 94(9) & 52.5 & 63.5(9) & 0.56* & 0.68 & 0.83\\
 \hline
\end{tabular}
\label{Tab:4}
\end{table}

\begin{table}[t!]
\caption{Measured and calculated activities for the zinc monitor. The measurements for analysis A were done at $t_{\mathrm meas}=$24h56m after EOI, measurements for analysis B were done at $t_{\mathrm meas}=$49h27m after EOI. The uncertainty of the FLUKA values corresponds to the statistical uncertainty. A starred value (*) indicates a C/E ratio which is outside the interval [0.6; 1.4].}
\centering
\footnotesize
\begin{tabular}{|c|c|c|c|c|c|c|c|c|c|}
\hline
{\bf Zinc} &  \multicolumn{2}{|c|}{Measured activity } & \multicolumn{2}{|c|}{Simulated activity } & \multicolumn{3}{|c|}{Comparison }\\
  & \multicolumn{2}{|c|}{[Bq]} & \multicolumn{2}{|c|}{[Bq]} &\multicolumn{3}{|c|}{C/E} \\
\hline
Reaction & Analysis & Result & MCNP6 & FLUKA & MCNP6 & FLUKA & C\textsubscript{M}/C\textsubscript{F} \\
\hline
\hline
 \textsuperscript{64}Zn(n,p)\textsuperscript{64}Cu & A & 1200(60) & 891 & 830(8) & 0.74  & 0.69 & 1.07 \\
 & B & 242(48) & 234 & 218(2) & 0.97 & 0.90 & 1.07\\
 \hline
  \textsuperscript{64}Zn(n,g)\textsuperscript{65}Zn & A & 3.2(7) & 1.7 & 2.26(2) & 0.53* & 0.71 & 0.75 \\
 & B & 2.8(4) & 1.7 & 2.25(2) & 0.61 & 0.80 & 0.76\\
 \hline
  \textsuperscript{68}Zn(n,g)\textsuperscript{69m}Zn & A & 15(1) & 6.3 & 8.3(4) & 0.42* & 0.55* & 0.76 \\
 & B & 6.3(6) & 1.8 & 2.41(12) & 0.28* & 0.38* & 0.75\\
 \hline
\end{tabular}
\label{Tab:5}
\end{table}
\Cref{Tab:2} gives the results for the multi-component monitor. The two analyses A and B were done with a time difference of 24 hours in the two different labs. Except for the \textsuperscript{58}Ni(n,np)\textsuperscript{57}Co reaction, the FLUKA code reproduces the experimental results reasonably well, with calculation-over-experiment (C/E) values between 0.77 and 1.20. MCNP6 gives results which are lower than the FLUKA ones by 20 to 40\%, except for the \textsuperscript{58}Co production, in which MCNP6 results give slightly better C/E ratios than FLUKA results. The lower values obtained with the MCNP6 code respect to the FLUKA results are consistent with the results on the source term in sec.~\ref{Sec:SourceTerm}, in which the MCNP6 results below 1 MeV are 40 to 50\% lower than the ones obtained with FLUKA.  
Both simulation codes consistently predict activities which are too low by factors 5 to 8 for the \textsuperscript{58}Ni(n,p)\textsuperscript{57}Co channel, while the experiments agree within errors (the long half-life of \textsuperscript{57}Co of 270 days allows to compare the measurements directly). An interesting fact is that the activity due to \textsuperscript{58}Co is larger for analysis B. This is due to the fact that there is a delayed production of \textsuperscript{58}Co due to the presence of the meta-stable state \textsuperscript{58m}Co which decays with a half-life of 9 hours to \textsuperscript{58}Co. This is taken care of in the simulations, but not in the extrapolation of the measurements back to the end of irradiation. For this reason it was decided to compare the simulations with the experimental results at the time of measurement. 

For the results of the indium monitor in~\cref{Tab:3}, we find that both codes predict a factor 3 to 4 less activity for the production of \textsuperscript{114m}In, but are remarkably close to the measurements for the channel \textsuperscript{115}In(n,n')\textsuperscript{115m}In. 

The results for the production of  \textsuperscript{117m}Sn with the tin monitor are given in~\cref{Tab:4}. The results with the FLUKA code give C/E ratios of 0.80 and 0.68, with the MCNP6 results being 17\% lower in both cases.

\Cref{Tab:5} gives the results for the zinc monitor. For the channel  \textsuperscript{64}Zn(n,p)\textsuperscript{64}Cu, MCNP6 gives results which have a C/E ratio of 0.74 for analysis A and 0.97 for analysis B, with FLUKA results being consistently 7\% lower. The activity due to the reaction   \textsuperscript{64}Zn(n,g)\textsuperscript{65}Zn is simulated by the FLUKA code with C/E ratios of 0.71 for analysis A and 0.80 for analysis B, while for the channel \textsuperscript{68}Zn(n,g)\textsuperscript{69m}Zn FLUKA predicts values significantly lower than the experimental results, namely a C/E ratio of 0.55 for analysis A and 0.38 for analysis B. For both reactions, MCNP6 predictions are about 25\% lower than the FLUKA predictions. 

In summary, for most of the reactions, the FLUKA code gives results with C/E ratios between 0.68 and 1.20, with the MCNP6 calculations giving in general results which are 10 to 40\% lower (with the exception of the \textsuperscript{64}Zn(n,p)\textsuperscript{64}Cu and the  \textsuperscript{58}Ni(n,p)\textsuperscript{58}Co reaction, for which MCNP6 results are 7 to 8\% higher than the FLUKA results).  

Both codes consistently give lower results for the production of \textsuperscript{114m}In and  the channel \textsuperscript{68}Zn(n,g)\textsuperscript{69m}Zn, with C/E ratios between 0.34 and 0.55 for FLUKA results and 0.24 and 0.42 for MCNP6 results. An especially large deviation between simulation and measured values is found for the reaction \textsuperscript{58}Ni(n,np)\textsuperscript{57}Co channel, with C/E ratios 0.16 and 0.18 for the FLUKA results and 0.12 and 0.14 for results obtained with MCNP6. Again, for these three channels, the MCNP6 results are lower than the FLUKA results by 20 to 30\%.

\subsection{Uncertainties in the calculations}

To estimate the uncertainties on the calculations, we need to address the different terms in \cref{Eq:1}. The primary sources of uncertainties in the simulations are the neutron flux rate at the sample position and the reaction cross sections. The neutron flux rate depends on the calculation of the source term (and therefore the underlying model for proton-induced neutron production in the water target), the proton beam current and to some extend the modeling of the target geometry. As mentioned in \cref{CalcSampAct}, the geometry of the system has been implemented with great care. The two geometric models differ only in minor details (see fig.~\ref{Fig:Geometry}), and contain both identical material compositions and densities. We therefore consider systematic effects from the geometry negligible when comparing the two simulations, and also they are thought to have a minor effect when comparing simulation results to the measurement, even given the fact that the influence of possible backscattering of neutrons from surrounding walls was not considered.  The neutron flux rate scales linearly with the proton beam current, and a deviation of the current from the nominal value of 25 $\upmu$A will reflect on the neutron flux rate. However, the beam parameters can be determined very well and their uncertainties are very small. The dominant effect comes from the  uncertainty of the source term - as can be seen from ~\cref{Fig:nRate} this can reach up to 50\% below energies of 1 MeV, and is certainly the reason why the MCNP6 results for most channels are significantly lower than the ones from FLUKA. A hint at the size of the uncertainty (which is energy dependent) is given by the C\textsubscript{M}/C\textsubscript{F}-values in \cref{Tab:2,Tab:3,Tab:4,Tab:5}. As can be seen, it can reach up to 40\% and more in some cases. 

\begin{figure}[th!]
\centering
\includegraphics[trim=95 100 55 95 ,clip, width=0.8\linewidth]{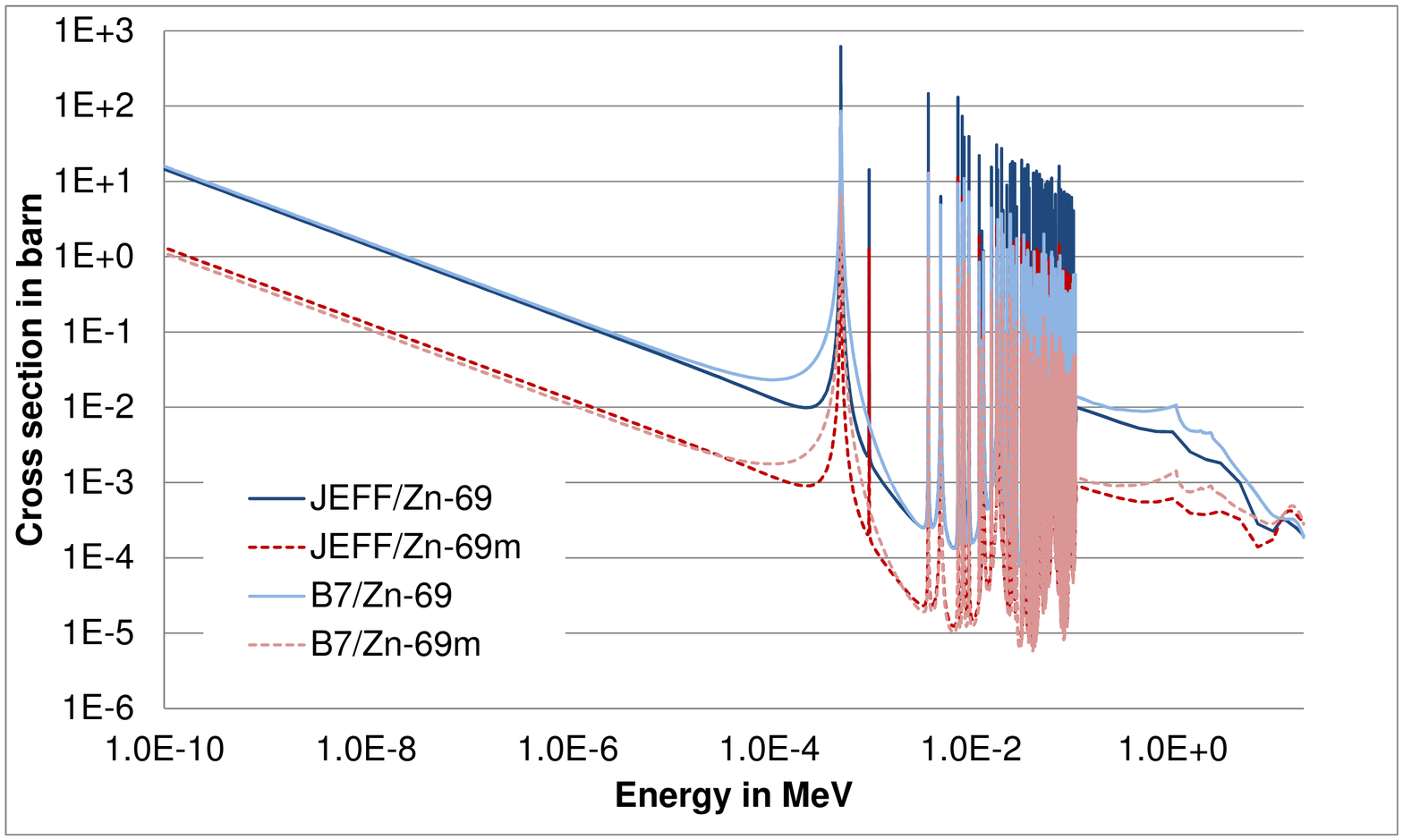}
\caption{Evaluated cross sections for the reaction \textsuperscript{68}FacureZn(n,g)\textsuperscript{69m}Zn as obtained with the JEFF3.1A and the ENDF/B-VII data libraries.}
\label{Fig:7}
\end{figure}

Fundamental for the determination of the reaction rates are the data libraries used in the evaluations. Different evaluated nuclear data libraries are available for the majority of nuclides at e.g. NEA~\cite{NEA}. For some nuclides only data calculated from theoretical models is available. Using MCNP6's capability to include different cross section data sets, we have studied the effects of different cross sections on the simulation calculations. The cross section data was generated with the use of the NJOY program. Among the libraries tested were ENDF/B-VII, JEFF3.1A, JENDL33, EAF2010 and ROSFOND2010. If branching fractions for isomeric states were given, these were considered. While for some of the considered reactions (like \textsuperscript{115}In(n,n')\textsuperscript{115m}In, \textsuperscript{117m}Sn production and \textsuperscript{197}Au(n,g)\textsuperscript{198}Au), the different libraries gave consistent results, in some cases discrepancies on the order of 30 to 40\% could be observed (\textsuperscript{114}In production and the \textsuperscript{68}Zn(n,g)\textsuperscript{69m}Zn reaction). As an example, in~\cref{Fig:7} the evaluated cross section for the \textsuperscript{68}Zn(n,g)\textsuperscript{69}Zn/\textsuperscript{69m}Zn reaction is shown obtained from two different data libraries (JEFF3.1A and the ENDF/B-VII). It can be seen that while for thermal energies, the cross sections are in good agreement, at high energies and around the first large resonance, the ENDF/B-VII data gives higher values, resulting in an activation result which is about 30\% higher than the one calculated with the JEFF3.1A data. For the remaining reactions, the differences in the activation results due to the use of different cross section data sets is on the order of 10\%. 

Another point is the influence of the energy group structure. While the FLUKA code uses a fixed 260 energy groups structure, one can have a larger number of groups for the cross section data sets generated with NJOPY for MCNP6. For the reaction \textsuperscript{58}Ni(n,p)\textsuperscript{58}Co activities were calculated based on ENDF/B-VII.1 data determined with a 260 and 640 energy group structure. One finds an effect of 10\% towards a better agreement with the experiments using the high-resolution structure. This shows a general problem in  the calculation of group-wise cross-sections. These may be underestimated in some cases, especially for threshold reactions. Generally, the flux decreases in the high energy region but cross-sections increase very strongly. Therefore it is very important to resolve the upper energy range well.   

The volumes of the monitors in the two sample stacks are not fully identical, the simulation results were calculated using the average volume of two monitors of the same material. While this effect is at maximum 1\% for the Indium, Zinc and the multicomponent monitor, it reaches about 5.5\% for the Tin monitors (and therefore the simulation results on the \textsuperscript{117m}Sn production). Of course, this is negligible compared to the potentially large uncertainties on the neutron flux rate and the cross section spectra.

After the irradiation, we were notified by the operator of the cyclotron that the irradiation had to be stopped for about 10 minutes due to a vacuum problem. Once the problem was fixed, irradiation resumed to complete the 50 minutes of irradiation time. Due to the fact that the lifetimes of the produced isotopes in the reactions in~\cref{Tab:1} are quite long,  we do not expect a large effect due to this interruption. An additional simulation with the FLUKA package using an irradiation time profile of 30 minutes of beam, followed by 10 minutes of no beam and finally additional 20 minutes of beam gave indeed no significant differences within the statistical uncertainties respect to the calculation with a full uninterrupted beam for 50 minutes. 

Finally, the values for material densities $\varrho$ and half-life times $\lambda$ were taken from the literature or are included in the simulation codes and the corresponding uncertainties are considered negligible for the present study.

\section{Conclusion and Outlook}
Inspired by the calculations of a shielding assessment for a new cyclotron bunker, investigations on the neutron source term for the \textsuperscript{18}F production at a IBA Cyclone 18/9 cyclotron were carried out using the Monte Carlo transport and reaction codes MCNP6 and FLUKA. It was found that below 1 MeV, the MCNP6 code gave a differential neutron rate which is smaller than the one by FLUKA by up to 50\%. The total neutron production yield for both codes was about 3 times larger than the value obtained from~\cite{IAEA} for the exclusive \textsuperscript{18}F production channel. To validate the results of the Monte Carlo codes, a more realistic model of the target geometry for the Cyclone 18/9 cyclotron was created with the two Monte Carlo codes which was used to calculate the activation of small monitor sample foils made of different metals and alloys during a typical run of \textsuperscript{18}F production. These results were then compared to the actual activation of the sample foils after a  \textsuperscript{18}F run which was obtained using gamma spectroscopy with HPGe detectors at two independent laboratories. In total, 11 reactions were investigated, with C/E ratios between 0.6 and 1.4 for most cases. As a general trend, results calculated using the MCNP6 codes were 20 to 40\% lower than the ones obtained with FLUKA. This may be (partially) explained by the fact that the source term obtained with MCNP6 is lower than the one from FLUKA below 1 MeV. For three reactions, the Monte Carlo simulations were consistently giving much lower results than the measured data (C/E values as low as 0.12 were observed). This was the case for the \textsuperscript{58}Ni(n,np)\textsuperscript{57}Co reaction, the production of \textsuperscript{114m}In and the reaction \textsuperscript{68}Zn(n,g)\textsuperscript{69m}Zn. For these three reactions, the uncertainties discussed can not accommodate the discrepancies, and it is most likely that the underlying cross section data for these reactions is responsible for the results, and eventually this document may help to improve the cross section data base in the Monte Carlo programs in the future. 

 Despite the uncertainties of the measurements, the obtained results show that the calculation of the neutron source terms with the help of the methods and models which are implemented in radiation transport and reaction codes like MCNP6 and FLUKA should work better for a proton beam of 18 MeV than a calculation based solely on the \textsuperscript{18}F yield. This is consistent with observations in~\cite{Carroll}, which reports significantly higher neutron yields for proton energies above 12 MeV for evaluations using a full ALICE-91 calculation respect to evaluations with tabulated yield values for the \textsuperscript{18}O(p,n)\textsuperscript{18}F reaction only. However, our observations seem to contradict the experimental results in~\cite{Mendez,Hagiwara} which find reasonable agreement with the yield of 
1.115$\times$10\textsuperscript{10} n/s for 1 $\upmu$A of proton current at 18 MeV energy obtained from~\cite{IAEA} for the ${}^{18}$O(p,n)${}^{18}$F channel.

In order to consolidate our results, further experiments will be planned  at the new cyclotron of the HZDR. In these experiments, both the target material and the proton energy will to be varied. The aim is to provide validated absolute neutron fluence spectra for shielding calculations at medical cyclotrons. 

\section*{Acknowledgement}
The authors would like to thank S. Bartel and M. K\"ohler of VKTA for providing the analysis of the sample monitors. We also like to thank S. Preusche and F. Hobitz of HZDR for assistance and support with the sample irradiation at the cyclotron.

\end{document}